\documentclass[aps,prl,reprint,showkeys,superscriptaddress,longbibliography,eprint]{revtex4-1}
\usepackage[utf8]{inputenc}
\usepackage[english]{babel}

\usepackage[T1]{fontenc}
\usepackage{systeme}
\usepackage{amsmath}
\usepackage{amssymb}
\usepackage{mleftright}
\usepackage[separate-uncertainty=true, multi-part-units=single]{siunitx}
\usepackage[inline,shortlabels]{enumitem}

\renewcommand\vec[1]{\boldsymbol{\mathrm{#1}}}

\usepackage{graphicx,grffile,subfigure}
\graphicspath{{../figures/}}
\newlength\figwidth
\usepackage{url,hyperref}
\usepackage[table,usenames,dvipsnames]{xcolor}
\hypersetup{colorlinks=true, linkcolor=BrickRed,
  urlcolor=blue!50!black, citecolor=blue!50!black, hypertexnames=false}
\usepackage[capitalize]{cleveref}
\crefrangelabelformat{equation}{(#3#1#4)$-$(#5#2#6)}

\newcommand\AdResS{AdResS}

\renewcommand\epsilon{\varepsilon}
\renewcommand\phi{\varphi}
\renewcommand\theta{\vartheta}
\renewcommand\rho{\varrho}
\renewcommand\leq{\leqslant}

\renewcommand\vec[1]{\boldsymbol{\mathrm{#1}}}

\newcommand\kB{k_{\text{B}}}
\DeclareMathOperator\tr{tr}

\begin{document}


\title{Open Systems out of Equilibrium: Theory and Simulation}

\newcommand\FUBaffiliation{\affiliation{Freie Universit\"{a}t Berlin, Institute of Mathematics, Arnimallee 6, 14195 Berlin, Germany}}

\author{Roya Ebrahimi Viand}
\FUBaffiliation
\author{Felix H\"ofling}
\FUBaffiliation
\affiliation{Zuse Institute Berlin, Takustr.\ 7, 14195 Berlin, Germany}
\author{Rupert Klein}
\FUBaffiliation
\author{Luigi Delle Site}
\email{luigi.dellesite@fu-berlin.de}
\FUBaffiliation

\begin{abstract} 
We consider the theoretical model of Bergmann and Lebowitz for open systems out of equilibrium and translate its principles in the adaptive resolution molecular dynamics technique (AdResS). We simulate Lennard-Jones fluids with open boundaries in a thermal gradient and find excellent agreement of the stationary responses with results obtained from the simulation of a larger, locally forced
closed system. The encouraging results pave the way for a computational treatment of open systems far from equilibrium framed in a well-established theoretical model that avoids possible numerical artifacts and physical misinterpretations.
\end{abstract}

\maketitle

Living matter is characterized by
a continuous microscopic transformation of physical and chemical entities and thus biological systems are predominantly out of equilibrium in nature. On the other hand, modern nanotechnology makes increasing use of out-of-equilibrium set-ups to produce new mechanisms of technological interest. For example long-standing problems of technological relevance with recent progresses in experiment and simulation include
the control of fluid flows through nanopores \cite{Bocquet:NM2020,Faucher:JPCC2019,Samin:PRL2017,Falk:NL2010}, gas storage in microporous hosts \cite{hoeft:jacs2015,peksa:mmm2015},
and the diffusion and permeability in random media \cite{Spanner:2016,Cho:2012,Scholz:2012,Gniewek:PRE2019}.
Thermal gradients, in particular, give rise to a variety of nonequilibrium phenomena of interest in current molecular nanoscience such as
the evaporation of liquids \cite{heinen:jcp2016,Tan:NC2019},
the thermomolecular orientation of nonpolar fluids \cite{romer:2012},
effects of thermo-phoresis \cite{duhr2006,maeda:2011} {and -osmosis \cite{ganti:prl2017}},
separation in liquid mixtures \cite{debu:2001,Roy:PRE2018},
diffusion of polymers in a solvent \cite{wurger:2009,lervik:2009},
heat transfer in protein--water interfaces \cite{lervik:2010},
and the polarization of water \cite{bremse:2008} to cite but a few.
In this context, molecular simulation can provide techniques that may both support our understanding of the detailed mechanisms responsible for the observed phenomena and guide the design of particular molecular systems with optimized nonequilibrium properties for technological applications. However, molecular simulation, and its most popular variant molecular dynamics (MD) in particular, faces two interconnected problems when dealing with molecular systems out of equilibrium. MD techniques were developed originally for equilibrium situations, whereas out of equilibrium simulations not only have a more complex mathematical description but also call for related more complex computational protocols.
Non-equilibrium MD (NEMD) techniques have been desired and developed by the MD community since 50 years
and various conceptual approaches and efficient techniques are available today \cite{booknoneq}. Most of these techniques rely on the simulation of systems with a fixed and typically large number of particles, thereby simultaneously representing the---generally limited---region of physical interest and the surrounding environment. In the present work, we describe instead a non-equilibrium extension of the adaptive resolution simulation (AdResS) approach, which was developed in the past decade \cite{adress1,annrev,physrep} for systems in equilibrium. The goal of the approach is the reduction of the computational complexity of a molecular simulation by focusing the main efforts, with full details and high accuracy, only on those regions in which the physics of interest is taking place. The surrounding environment is, in contrast, simplified to just the essential degrees of freedom required to avoid undue losses of physical information relevant to the region of interest. The region of high resolution is thus effectively reduced to an open sub-domain of the total system that exchanges energy and particles with its environment, the latter playing the role of a thermodynamic reservoir at prescribed macroscopic conditions \cite{prx}. The new conceptual challenge in developing a non-equilibrium extension to this approach is the proper representation of the statistical mechanics of an open system that is out of equilibrium with its given surroundings. The aim of the present paper is to show that the model of Bergmann and Lebowitz (BL) for open systems far from equilibrium \cite{leb1,leb2} provides a theoretical framework that justifies the use of the AdResS approach as a simulation protocol for nonequilibrium situations, here specifically applied to a liquid with open boundaries in a thermal gradient.

\paragraph*{Adaptive resolution technique.}

\begin{figure}
\centering
\includegraphics[width=3.4in]{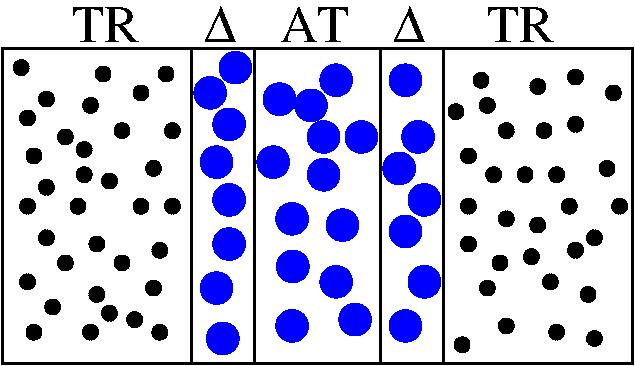}
\caption{Schematic illustration of the typical AdResS set-up.
The atomistically resolved region AT is the region of interest, in which particles evolve under Hamiltonian dynamics without artificial forcing.
The AT region is interfaced with the $\Delta$ regions where molecules have atomistic resolution and are, in addition, subject to a thermostat and to the one-particle ``thermodynamic force'', which assures the proper equilibrium in the AT region. Each $\Delta$ region is interfaced on the other side with a reservoir of tracers, i.e., with non-interacting particles whose thermal state is controlled by the same thermostat and thermodynamic force that act in the $\Delta$ region. Extreme repulsions between tracer particles whose centers are unphysically close upon entering the $\Delta$ region are controlled by capping the forces on these particles at a certain threshold.}
\label{cartoon1}
\end{figure}

AdResS in its latest version \cite{jcpabrupt,advtsres} consists of a simulation set-up where the space is divided in two parts, a region of interest where molecules have an atomistic resolution (AT), and a larger region (TR), with the role of a reservoir, where molecules are represented by non-interacting point particles (tracers) thermalized at a prescribed temperature by an imposed thermostat. At the interface between these two regions lies a region ($\Delta$) where molecules have atomistic resolution and are subject to the action of a thermostat and a one-particle force, ${\vec F}_{\text{th}}(\vec r)$, named thermodynamic force (see also the pictorial representation in \cref{cartoon1}).
Here, we restrict to set-ups with planar interfaces for simplicity; then ${\vec F}_{\text{th}}(\vec r)$ depends only on the position $x$ along the interface normal and consists only of the vector component along the $x$ direction.
In an equilibrium set-up, the latter enforces the expected thermodynamic, structural, and dynamic equilibrium properties within the atomistic subdomain, without noticeable artifacts. The derivation of the method is based on the equality in equilibrium of the grand potential of the atomistic region and the rest of the system \cite{prl2012}. ${\vec F}_{\text{th}}(x)$, in essence, induces a local balancing of the pressure, $p(x)$, to the fully atomistic value of reference.
Related studies \cite{prx,jcpanimesh} have shown that the work due to ${\vec F}_{\text{th}}(x)$ balances, together with the work of the thermostat, the difference in chemical potential between the atomistic region and the rest of the system.
For practical applications ${\vec F}_{\text{th}}(x)$ is calculated via an iterative procedure with the aim of eliminating in $\Delta$ local variations of the density with respect to the value at the desired thermodynamic state point as it would be obtained from a full atomistic simulation \cite{prl2012,softmatt,physrep}.
The converged function, ${\vec F}_{\text{th}}(x)$, is then employed in subsequent production runs without recalibration.

At the technical level, in the set-up of \cref{cartoon1}, molecules
of the AT region interact through atomistic force fields among themselves and with molecules in $\Delta$, whereas there is no direct interaction with the tracers.
The two contributions yield the potential energies $U_{\text{AT}-\text{AT}}$ and $U_{\text{AT}-\Delta}$, respectively.
Second, the thermodynamic force derives from a potential,
${\vec F}_{\text{th}}(\vec r) = - \nabla \phi_\text{th}(\vec r)$,
with the convention that $\phi_\text{th} = 0$ in the AT region \cite{advtsres}.
The total potential energy of the system is thus
$
  U_\mathrm{tot}=
  U_{\text{AT}-\text{AT}} + U_{\text{AT}-\Delta}
    + \sum_{j\in \Delta \cup \mathrm{TR}} \phi_\text{th}({\vec r_j}),
$
where $\vec r_j$ denotes the position of particle~$j$.
A relevant conceptual step in the elaboration of the method
reported above is the mapping of the AdResS protocol onto a well established
theoretical framework for the statistical mechanics of an open system
\cite{njp,preliouv,physrep,softmatt}. In fact, all the principles of the AdResS
protocol have been put in direct relation with the principles of the stochastic model of open systems developed by Bergmann and Lebowitz \cite{leb1,leb2}.

\paragraph*{Bergmann-Lebowitz model of open system}

In the BL model the open system is embedded in an environment of one or more reservoirs ($r=1,\dots,m$) with which the system exchanges energy and particles. The coupling between the system and reservoir $r$ consists of an impulsive interaction at discrete
points in time, which is mathematically formalized by a suitable kernel $K^r_{nn'}(X'_{n'},X_{n})$. This represents the probability per unit time that, due to the interaction, the $n$-particle open system with phase space configuration $X_{n}$ makes a transition to $n'$ particles with phase space configuration $X'_{n'}$.
In the evolution equation for the $n$-particle phase-space probability $f_n(X_{n},t)$,

\begin{equation}
  \frac{\partial f_n(X_{n},t)}{\partial t}=\{f_n(X_{n},t),H_n(X_{n})\} + I_n(X_{n}, f(t)) \,,
  \label{liouvext}
\end{equation}
the last term
\begin{widetext}
\begin{equation}
 I_n(X_{n}, f(t)) := \sum_{r=1}^m\sum_{n'=0}^{\infty}\int dX'_{n'} \bigl[
  K^{r}_{nn'}(X_{n},X'_{n'})f_{n'}(X'_{n'},t)
  - K^{r}_{n'n}(X'_{n'},X_{n})f_n(X_{n},t)
 \bigr],
  \label{interaction}
\end{equation}
\end{widetext}
depends on the full hierarchy $f(t) = \{f_{n'}(\cdot, t)\}_{n'=0,1,\dots}$ at time $t$ and
expresses the interaction between the system and the $m$ reservoirs.
\Cref{liouvext} is the equivalent of Liouville's equation for an open system in contact with several reservoirs. The development of a systematic procedure for deriving an analytic form of $K^{r}_{nn'}(X_{n},X'_{n'})$ for complex many-particle systems represents, until now, a formidable task. For this reason, in molecular simulations one can only design algorithms which mimic, as close as possible, the expected action of the kernel, $K^{r}_{nn'}(X_{n},X'_{n'})$, without knowing its exact analytic expression. Along these lines we now proceed with a discussion of the analogy between the AdResS protocol and the BL model.

\paragraph*{Analogy of AdResS and the BL model.}

The AT region in the AdResS scheme can be interpreted as an open system in the sense of the Bergmann--Lebowitz model under the approximation that the reservoir (TR and $\Delta$ region) is large enough  and that
$U_{\text{AT}-\text{AT}} \gg U_{\text{AT}-\Delta}$,
i.e., that the interaction energy between the particles in the atomistic region and the particles in $\Delta$ is negligible. Furthermore, the action of the transition kernel in the BL model, in AdResS corresponds to (i) the force which particles in $\Delta$ impose on particles in AT and (ii) to the exchange of particles between AT and $\Delta$. The former impose changes of momentum and energy of the particles residing in AT, the latter allows for changes of the number of particles inside the AT region. The combined action of the thermodynamic force and the thermostat in $\Delta$ guarantees that the statistics of the $\Delta$-particles is maintained at the desired reservoir state \cite{njp,preliouv,physrep,softmatt}.
 
In its general form, the BL model allows for the instantaneous exchange of an arbitrary number of particles through the action of the stochastic exchange kernels. At the same time the system state can undergo arbitrary changes in phase space as well. In contrast, in the situation we consider in AdResS, i.e., that of a dynamically evolving open system, changes of the particle number larger than one and state changes that involve particles far away from the system boundaries are extremely unlikely. Therefore, a specification of the BL kernel to this situation would call for setting $K_{nn'}=0$ whenever $|n-n'|>1$. Similarly, for substantial transition probabilities $K_{nn'}(X_{n},X_{n'})$, $X_{n}$ and $X_{n'}$ should be nearly identical for all particles far away from the boundaries \cite{liouv-prr,prog-rep-cont}. Conceptually, the AdResS approach implements these constraints in that (i) number changes are induced by a dynamical process continuous in time for which simultaneous crossing of the boundaries by more than one particle is extremely unlikely and (ii) particles entering or leaving the open system would at this instant influence only their immediate surroundings but not the entire system domain. In this sense, one way to interpret the AdResS set-up is as a dynamic-like approximation of the BL stochastic process or vice versa. Numerical simulations {showed} that indeed under such a framework AdResS follows the grand canonical behaviour {predicted within} the BL model for equilibrium (see e.g.\ Refs.~\cite{njp,advtsres,physres}).

\begin{figure}
\centering
\includegraphics[width=3.4in]{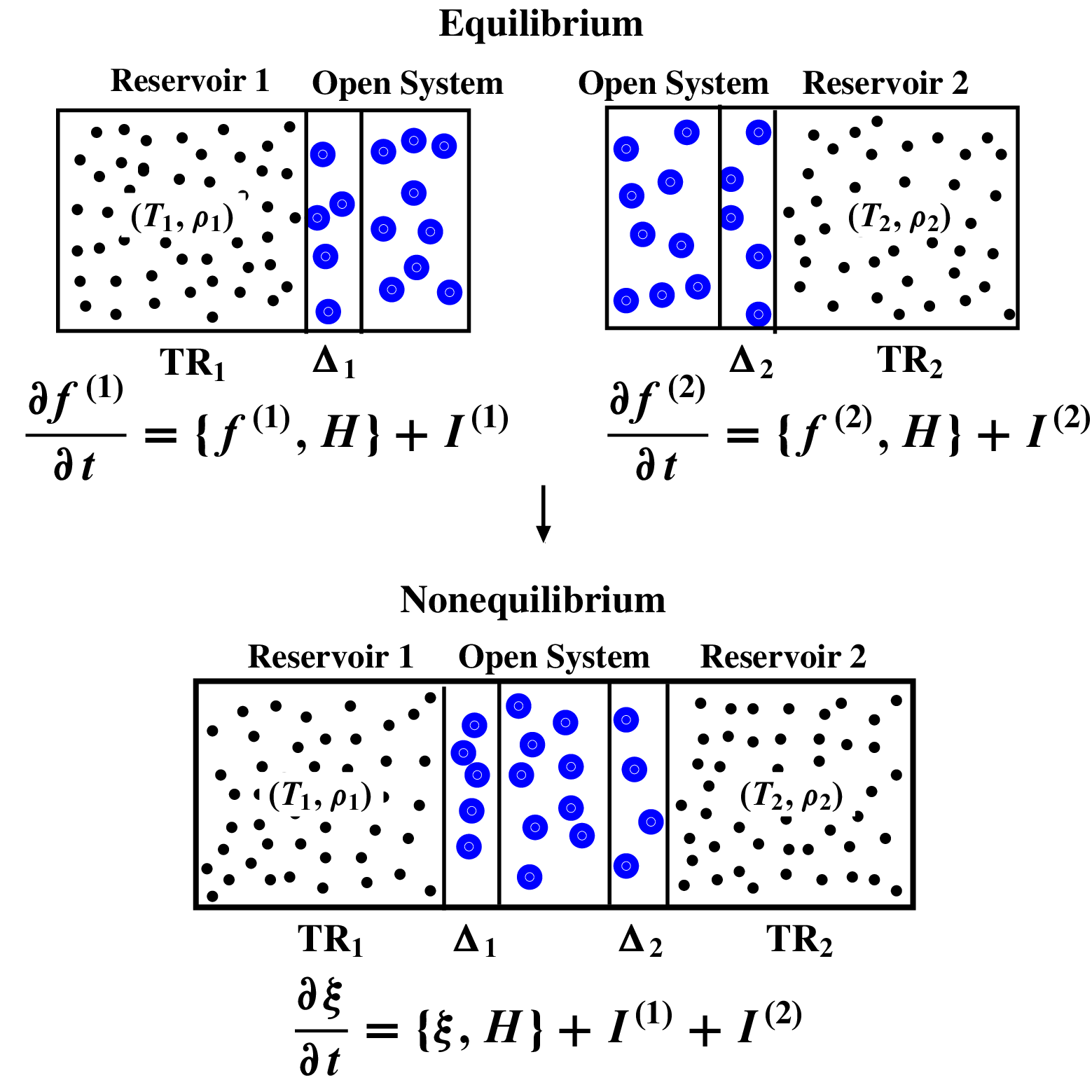}
\caption{Schematic illustration of the meaning of a thermal gradient in the BL model and its corresponding mapping into the AdResS set-up. When interacting only with one reservoir the system equilibrates at the thermodynamic condition {of the reservoir}. In the BL model {[\cref{liouvext,interaction}]}, this is equivalent to defining the transition term $I^{(i)}$ {for system $i=1$ or $2$}, while in AdResS it means that the $\Delta_i$ {and $\text{TR}_i$ regions} are {subject to} a thermostat with target temperature $T_{i}$ and the corresponding thermodynamic force, ${\vec F}^{(i)}_{\text{th}}(x)$. Once the system is in contact with two different {reservoirs}, then in the BL model one has the combined effect of $I^{(1)}+I^{(2)}$, which, in the AdResS set-up, translates into
{a region $\Delta_{1} \cup \text{TR}_1$ forced at temperature $T_{1}$ and by ${\vec F}^{(1)}_{\text{th}}(x)$ and a region $\Delta_{2} \cup \text{TR}_2$ with parameters $T_{2}$ and ${\vec F}^{(2)}_{\text{th}}(x)$.}}
\label{cartoon2}
\end{figure}

\paragraph*{Nonequilibrium of an open system.}

Bergmann and Lebowitz \cite{leb1} pointed out that, according to their model, a system connected to two (or more) reservoirs with different thermodynamic conditions, e.g., different temperatures, in the stationary state will have  heat (and {mass}) currents flowing through the system. Formally, such currents are produced by the interaction terms $I_n$ ($n=0,1,\dots$) of the extended Liouville equation \eqref{liouvext}, with $I_n$ defined in \cref{interaction}.
For the AdResS simulation the equivalent effect, according to the analogy with the BL model, {is produced by coupling the region AT of interest to two reservoirs that are at different thermodynamic state points of the fluid, specified for example by their temperature and density, $(T_1, \rho_1)$ and $(T_2, \rho_2)$, respectively (\cref{cartoon2}).
To this end, one needs to pre-compute the thermodynamic force $\vec F_\text{th}^{(1)}(\vec r)$ at the state point $(T_1, \rho_1)$ and, separately, $\vec F_\text{th}^{(2)}(\vec r)$ at the other state point $(T_2, \rho_2)$; the result may be stored in a ``dictionary'' mapping pairs $(T_i, \rho_i)$ to $\vec F_\text{th}^{(i)}(\vec r)$ for later reference.
The nonequilibrium AdResS set-up is then realized by employing $\vec F_\text{th}^{(1)}(\vec r)$ together with a thermostat at the temperature $T_1$ in the $\Delta_1$ and $\text{TR}_1$ regions and correspondingly for the second reservoir using $\vec F_\text{th}^{(2)}(\vec r)$ and a thermostat at $T_2$.}

To demonstrate that this protocol will indeed approximate the behavior of a large molecular system with spatially separated {thermodynamic} forcings, we {performed numerical experiments on Lennard-Jones (LJ) fluids} that are subject to a temperature gradient {(see below)}.
{If a set of relevant physical observables, computed from the AT region of the set-up only,}
agrees with the results of a full atomistic simulation, {it is corroborated} that the BL model with localized exchange kernels provides a solid theoretical reference for AdResS simulations {far from} equilibrium
and {we can conclude that} the combination of the BL model with the AdResS protocol provides a promising basis for further development in the theory and simulation of open systems, in and out of equilibrium \cite{liouv-prr,prog-rep-cont}.

\begin{figure*}
\centering
\includegraphics[width=\linewidth]{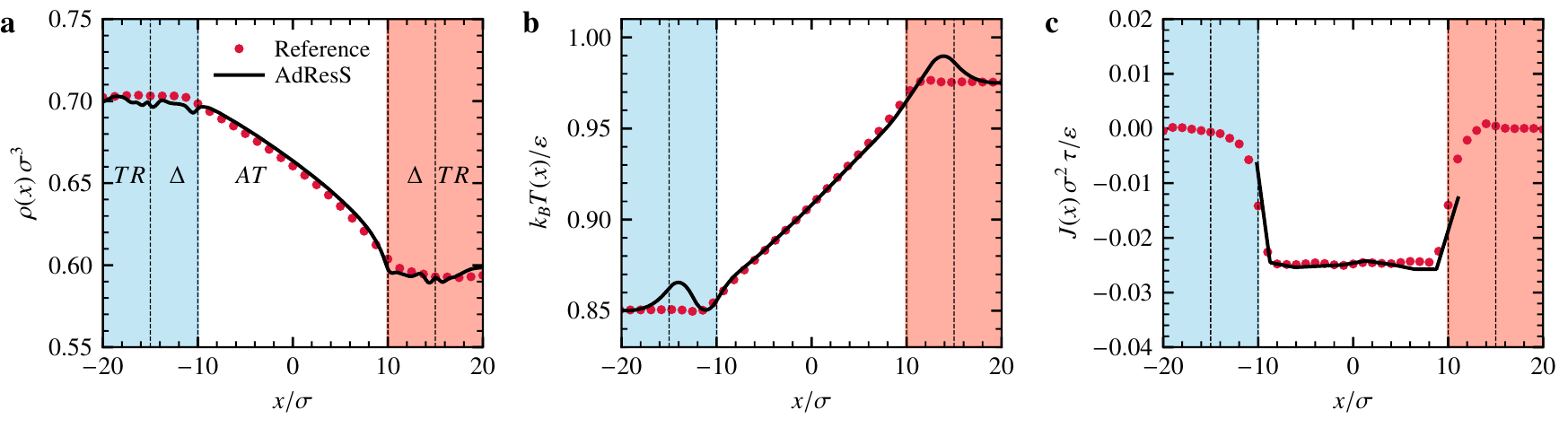}
\caption{Results of open-system simulations for a Lennard-Jones (LJ) liquid in a thermal gradient obtained from the nonequilibrium AdReS set-up (solid lines, \cref{cartoon2}) and from a full atomistic simulation serving as reference (dots).
Both set-ups share the same geometry of the simulation box, in particular, the same size of the region of interest (AT) and the same box length $120\,\sigma$ along the $x$-axis; parts of the tracer regions (TR) have been omitted for clarity.
The units of energy, length, and time are given by the LJ parameters $\epsilon$, $\sigma$, and $\tau$, respectively. 
The panels display spatial profiles of
(a)~the particle number density $\rho(x)$,
(b)~the temperature $T(x)$, and
(c)~the heat flux $J(x)$,
as stationary responses to the nonequilibrium forcing.
}
\label{sys}
\end{figure*}

\paragraph*{Nonequilibrium simulations.}

{Motivated by the study of \citet{ciccgrad}, we simulated the stationary response of a LJ fluid to an applied temperature gradient using the nonequilibrium AdResS set-up described above (\cref{cartoon2}).
The stationary state of the analogous full atomistic simulation is characterized by mechanical equilibrium, in which pressure gradients are balanced.
We have accounted for this fact by choosing the reservoir states 1 and 2 along an isobar of the fluid:
for given temperatures $T_1$ and $T_2$, we determined the densities $\rho_1$ and $\rho_2$ that yield equal pressures, $p(T_1, \rho_1) = p(T_2, \rho_2)$, according to the equation of state.
For the initial equilibrium state, we used a state point in the liquid phase with number density
$\overline\rho = (\rho_1 + \rho_2) / 2 = 0.65 \,\sigma^{-3}$ and temperature
$\overline T = (T_1 + T_2) / 2 \approx 0.91 \,\epsilon/\kB$;
the symbols $\sigma$ and $\epsilon$ serve as units of length and energy and refer to the parameters of the LJ potential, which was truncated at $r_c = 2.5 \,\sigma$
(see also Supplemental Material 
).
A temperature difference of $\Delta T = T_2 - T_1 = 0.125 \, \epsilon/\kB$ was then imposed symmetrically.
To this end, the AdResS set-up was first equilibrated at $(\overline T, \overline\rho)$ using the same, suitably calibrated reservoir parameters on both sides, which yields a uniform density and temperature across the whole set-up comprised of AT, $\Delta$, and TR regions.
Then, the states of the two reservoirs were switched to $(T_1, \rho_1)$ and $(T_2, \rho_2)$ by (i) replacing the thermodynamic force of the equilibrium set-up by the forces $\vec F_\text{th}^{(1)}(\vec r)$ and $\vec F_\text{th}^{(2)}(\vec r)$, respectively (see above), and by (ii) changing the thermostat temperatures to $T_1$ and $T_2$.}

{Both the nonequilibrium AdResS set-up and the full atomistic simulation, serving as a benchmark for reference, were implemented in the simulation framework ``HAL's MD package'' \cite{colberg2011,*HALMD}, which features large systems due to massive parallelization provided by high-end graphics processors and, concomitantly, excellent numerical long-time stability of Hamiltonian dynamics; the technical details can be found in the Supplemental Material.}
We have verified that the AT regions of the equilibrium set-ups of AdResS for the calculation of ${\vec F}^{1}_{\text{th}}(\vec r)$ and ${\vec F}^{2}_{\text{th}}(\vec r)$, satisfy the conditions of
open system (see Supplemental section).
The resulting thermodynamic forces can be found along with their potentials in Supplemental Fig.~S2.
{Nonequilibrium averages were computed as in Ref.~\citenum{ciccgrad} using} the D-NEMD approach developed by \citet{dnemd}.
{It} consists of running MD simulations for an equilibrium (or, at least, stationary) state of reference, e.g., a fluid at temperature $\overline T$.
Along the obtained system trajectory a series of uncorrelated samples is selected, {from which an ensemble of nonequilibrium trajectories is branched off by} taking these samples as initial configurations of the NEMD simulations with, e.g., an imprinted thermal gradient.

{Following \citet{ciccgrad} we have considered the observables density, temperature, and heat flux and calculated their spatial profiles $\rho(x)$, $T(x)$, and $J(x)$, respectively,
which emerge as the stationary response to the nonequilibrium forcing by the reservoirs.
The absence of many-body interactions permits the calculation of the heat flux
in a slab-like control volume $\Omega$ centered at position $x^*$ as \cite{ciccgrad,heat}
$
  J(x^*)= V_\Omega^{-1}\sum_{j\in \Omega} (e_j + \tr \vec\sigma_j) \vec v_j \,;
$
therein, $e_j$, $\vec\sigma_j$, and $\vec v_j$ denote the total energy, the potential part of the stress tensor, and the  velocity of the $j$-th particle, respectively, and $V_\Omega$ is the volume of $\Omega$.}
Within the AT region, {where the molecular dynamics evolves freely without a thermostat nor a thermodynamic force,} all three profiles obtained from the AdResS set-up show {excellent} agreement with the results of the corresponding full atomistic simulation (\cref{sys}).
In the $\Delta$ region, the system is artificially forced so that an agreement
with the reference simulation is neither required nor expected; {most importantly, the small deviations of the temperature profile $T(x)$ near the $\Delta$/TR boundary relax within the $\Delta$ region.}

{As a second test, we performed simulations along the same lines for a supercritical LJ fluid at moderate density $\overline\rho = 0.3\,\sigma^{-3}$ and elevated temperature $\overline T = 1.5 \,\kB/\epsilon$, well above the liquid--vapour critical point, with a symmetric temperature difference of $\Delta T = 0.2 \,\kB/\epsilon$.
This fluid is more compressible and exhibits a six-fold higher pressure than the above liquid.
The obtained profiles are reported in Supplemental Fig.~S4 and show the same high degree of agreement as found for the liquid case.}

\paragraph*{Conclusions.}

We introduce a generic framework for MD of open systems out of equilibrium and numerically treat the case of an open boundary LJ fluid in a thermal gradient. We have shown that an AdResS numerical set-up that follows a tight analogy to the BL stochastic model of open system out of equilibrium {can accurately reproduce} the {full} atomistic simulations of large systems serving as benchmark reference. Such results allow one to employ AdResS and the BL model with localized exchange kernels as a prototype theoretical and numerical model of reference in the development and application of open system approaches in molecular simulation. The computational advantage of such an approach is very relevant for simulation studies with very complex molecular environments. In fact it can simplify a complex molecular system by defining a region of observation where the process of interest takes place and a generic environment described by the transition kernel in the BL model and correspondingly by the thermostat and the thermodynamic force in AdResS. An appealing perspective offered by the method concerns its possible use in particle-continuum approaches and the possibility of performing simulations involving mass flow which so far have required problem-specific, tailored solutions not transferable to other situations (see Ref.\citenum{prog-rep-cont} and references therein).


\section*{Acknowledgments} 
This research has been funded by Deutsche Forschungsgemeinschaft (DFG) through grant CRC 1114 ``Scaling Cascade in Complex Systems,'' Project Number 235221301, Project C01 ``Adaptive coupling of scales in molecular dynamics and beyond to fluid dynamics.''


\bibliography{paper}



\clearpage
\onecolumngrid

\section*{Supplementary Material}

\twocolumngrid


\makeatletter\floats@sw{
\renewcommand{\theequation}{S\arabic{equation}}
\renewcommand{\thefigure}{S\arabic{figure}}
\renewcommand{\thetable}{S\arabic{table}}
\setcounter{equation}{0}
\setcounter{figure}{0}
\setcounter{table}{0}
}{}
\makeatother

\section{Simulation details}

\subsection{Physical parameters}

The simulated Lennard-Jones (LJ) fluids consist of point particles of mass $m$ interacting via the smoothly truncated and shifted pair potential
$
 U(r) = [U_\text{LJ}(r) - U_\text{LJ}(r_c)] f((r - r_c)/h)
$
for $r \leq r_c$, and $U(r) = 0$ otherwise, with
$
 U_\text{LJ} = 4 \epsilon \bigl[ (r/\sigma)^{-12} - r/\sigma)^{-6} \bigr] ,
$
the cutoff radius $r_c = 2.5\sigma$, the truncation function
$f(x) = x^4/(1+x^4)$, and $h=0.005\sigma$ \cite{zausch2010,colberg2011,Roy:JCP2016}.
The parameters $\epsilon$ and $\sigma$ serve as units for energy and length,
$\tau=\sqrt{m \sigma^2/\epsilon}$ defines the unit of time,
and dimensionless quantities are given by $\rho^* = \rho\sigma^3$ and $T^* = \kB T/\epsilon$.
The tracer particles in the AdResS set-up do not interact with each other and not with the LJ particles in the AT and $\Delta$ regions, which is achieved by setting $\epsilon=0$ for the interactions involving tracers.

The simulation results reported in Fig.~3 of the main text were obtained for two liquid states along the same isobar. The first point was chosen at temperature $T_2^* = 0.975$ and density $\rho_2^* = 0.5987$,
right at the liquid--vapour coexistence line \cite{Vrabec:MP2006}, yielding a (reduced) pressure of $p^* := p \sigma^3/\epsilon = 0.052$.
For the second point, we used the lower temperature $T_1^* = 0.850$ and, from a small sequence of simulations, determined the density $\rho_1^* = 0.7047$ at which the two liquids have the same pressure,
$p(T_1, \rho_1) = p(T_2, \rho_2)$.
An accurate equation of state and the phase diagram for the truncated LJ potential can be found in Ref.~\cite{Heier:MP2018}.

A second set of simulations was carried out for LJ fluids in the supercritical regime, at moderate density and well above the liquid--vapour critical temperature ($T^*_c \approx 1.08$).
Specifically, we have chosen the two state points $(T_1^*, \rho_1^*) = (1.40, 0.350)$ and
$(T_1^*, \rho_1^*) = (1.60, 0.248)$, corresponding to a pressure of $p^*=0.32$. The corresponding results are reported in Fig.~\ref{dense-gas}

\subsection{Implementation and algorithmic parameters}


For AdResS set-ups in its most recent form as employed here \cite{advtsres}, the following capabilities are needed beyond standard MD techniques:
\begin{enumerate}[label={(\roman*)}]
 \item partitioning of the the simulation domain into the regions AT, $\Delta$, and TR and unions thereof,
 \item a stochastic thermostat acting on such subdomains,
 \item a mechanism for the change of resolution that flips molecules (here: LJ beads) into tracers and back,
 \item the thermodynamic force calculated from an external, one-particle potential parametrized on a grid,
 \item the capping of excessively large forces between molecules in the $\Delta$ region.
\end{enumerate}
We have implemented these requirements into the simulation framework ``HAL's MD package'' \cite{colberg2011,*HALMD},
which has proven as an efficient and accurate tool for large-scale MD studies of the dynamics in liquids \cite{Roy:JCP2016,Straube:CP2020}.
Data sets for particle trajectories and time series of thermodynamic observables were stored in the binary and compressed hierarchical file format H5MD \cite{h5md}.


\begin{figure}[b]
\centering
\centerline{\includegraphics[width=3.4in]{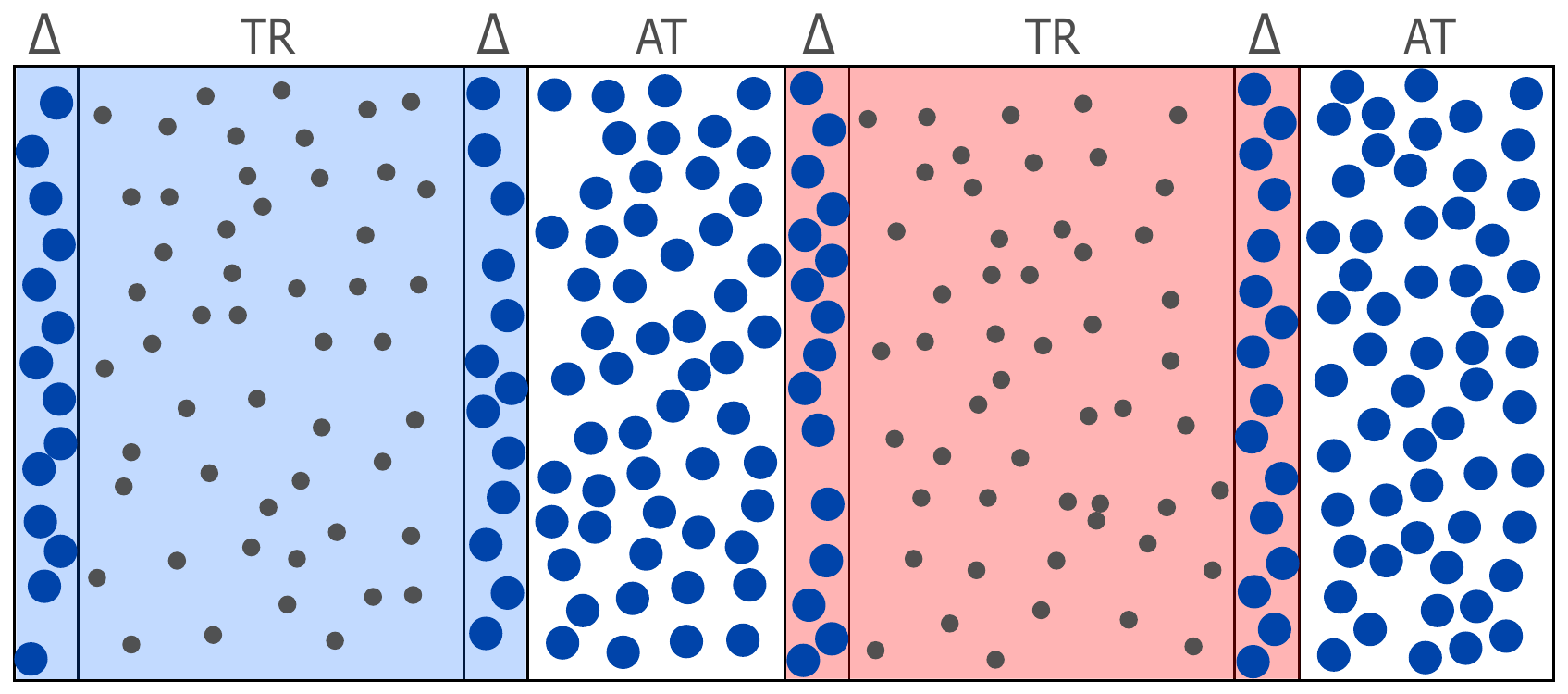}}
\caption{Extending the nonequilibrium AdResS set-up (Fig.~2 of the main text) by its mirror set-up facilitates periodic boundary conditions on all faces of the simulation box. The two mirror-symmetric AT regions yield independent samples of the observables.}
\label{mirrored-setup}
\end{figure}

In all nonequilibrium simulations performed, we used a cuboid simulation box of size $120\sigma \times 20\sigma \times 20\sigma$, where the first dimension refers to the direction along which molecules change their resolution, denoted as $x$-axis.
Periodic boundary conditions were applied on all faces of the box, and to this end, the set-up was extended by its mirror image (\cref{mirrored-setup}).
The final set-up contained two independent AT regions (the regions of interest), which separately entered the ensemble average for the calculation of the observed profiles.
For the AdResS set-up, each AT region of width $20\sigma$ was sandwiched along the $x$-axis by two transition regions $\Delta$ of width $5\sigma$, the remaining space was divided in two equally sized tracer regions (TR) of width $30 \sigma$ each.

The total number of particles in the system (LJ beads and tracers)
was such that it matches the average density $\overline\rho = (\rho_1 + \rho_2) / 2$ of the corresponding nonequilibrium states, e.g., 31,282 particles for the liquid case,
and the same number of particles was used in the corresponding full atomistic simulation.
Non-equilibrium trajectories over a duration of $15{,}000\tau$ were generated with the velocity Verlet integrator with timestep $0.002\tau$.
The first quarter of each trajectory ($3{,}750\tau$) was discarded for the data accumulation of the stationary profiles.

Particles in the AT region are subject to the unmodified Hamiltonian dynamics due to the atomistic force fields.
The $\Delta$ and TR regions were thermalized with the Andersen thermostat \cite{andersen},
with the update rate set to $50\tau^{-1}$, i.e., a particle's velocity is re-sampled from the Maxwell--Boltzmann distribution every 10 integration steps on average.
The choice of the rate influences the peaks of the temperature profiles in the $\Delta$ region (Fig.~3b of the main text), which are diminished by a tighter coupling of the thermostat to the system.


The thermodynamic force $F_\text{th}(x)$ was parametrized on a uniform grid along the $x$-axis with a knot spacing of $0.25\sigma$ using an interpolation in terms of a cubic Hermite spline for the potential $\phi_\text{th}(x)$.
The total force on a particle was capped at a threshold of $500\,\epsilon/\sigma$ for each Cartesian vector component while preserving the sign of the component.
After every integration step, LJ beads whose centers entered the TR region were changed into tracers, and tracers that entered the $\Delta$ region where changed into LJ beads.

\subsection{Nonequilibrium simulation protocol}

For the nonequilibrium simulations, we made use of the D-NEMD technique \cite{dnemd,ciccgrad} to generate an ensemble of trajectories from uncorrelated initial conditions. Both the AdResS and the full atomistic reference simulations followed the same protocol:
\begin{enumerate}[label={(\roman*)}]
 \item Perform one equilibrium simulation at temperature $\overline T=(T_1 + T_2)/2$ and density $\overline\rho = (\rho_1 + \rho_2)/2$. It yields the trajectory of a homogeneous fluid along which configurations are sampled every $40\tau$ after an initial equilibration phase of $2{,}000\tau$.

 \item Start non-equilibrium simulations from these samples. In the AdResS set-up, the two reservoirs for the equilibrium set-up at $(\overline T, \overline \rho)$ are replaced by one reservoir at $(T_1, \rho_1)$ and one at $(T_2, \rho_2)$, i.e., the thermostat temperature and the parameters of $F_\text{th}(x)$ are changed.
 For the full atomistic reference, only the thermostat is modified.
\end{enumerate}
In the case of AdResS, the thermodynamic force was pre-computed for the three state points used.
Note that the full atomistic simulations are not needed for AdResS simulations according to this protocol, they served as a benchmark reference merely.

\subsection{Observables}

For the calculation of the spatial profiles of thermodynamic observables, the simulation box was partitioned into slab-like control volumes $\Omega_k$ of width $2.5\sigma$ and volume $V_\Omega$ along the $x$-axis ($k=1,\dots, 48$).
The temperature $T(x_k)$ at the position $x_k$ in the center of $\Omega_k$ follows from the kinetic energy of the particles in $\Omega_k$.
The heat flux was obtained as
$
  J(x_k) = V_\Omega^{-1}\sum_{i\in \Omega_k} (e_i + \tr \vec\sigma_i) \vec v_i
$
\cite{ciccgrad,heat} with $\vec v_i$ denoting the velocity of the $i$-th particle, $e_i$ the sum of its kinetic and potential energies, and
$
  \tr \vec \sigma_i = -\frac{1}{2} \sum_{j \neq i} r_{ij} \, U'(r_{ij})
$
the trace of the potential contribution of particle $i$ to the stress tensor;
$r_{ij} = |\vec r_i - \vec r_j|$ is the distance of separation between particles $i$ and $j$.

After giving ample time for relaxation of the non-equilibrium setup ($3{,}750\tau$), the profiles were computed as time averages over samples taken every $0.3\tau$.
The data shown in Figs.\ 3 (main text) and \ref{dense-gas} are averages over time, the 4 independent nonequilibrium trajectories, and the two independent AT regions in the simulation box.

\section{\protect\AdResS{} simulations in equilibrium}

In this section, we describe the procedure to calculate the thermodynamic force used in the AdResS set-ups
and we give numerical evidence that the AT region of the set-up properly represents a grand canonical open system.

\subsection{Computation of the thermodynamic force}

\begin{figure}
\centering
\centerline{\includegraphics[width=3.4in]{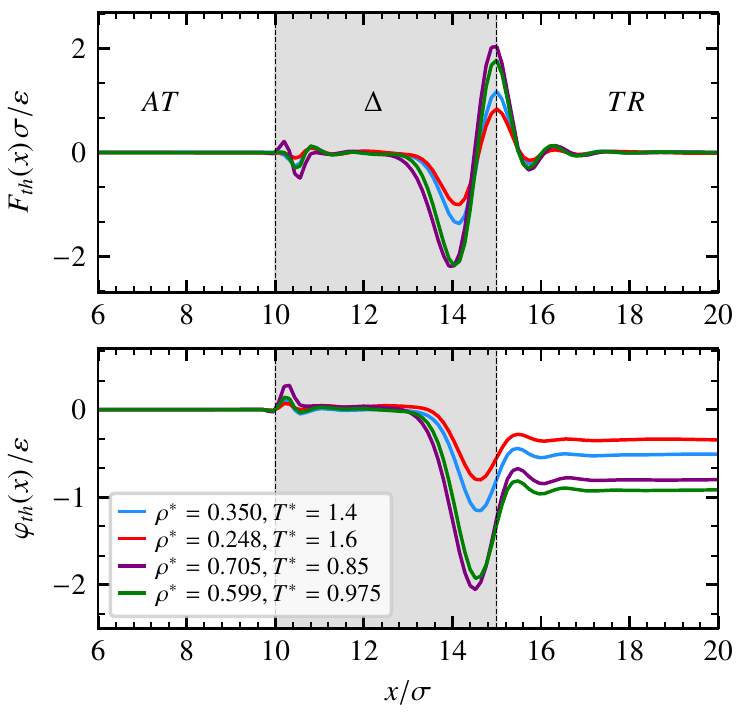}}
\caption{The thermodynamic forces $F_\text{th}(x)$ (top) and their corresponding potentials $\phi_\text{th}(x)$ (bottom) along the $x$-axis for the four reservoir states used in the nonequilibrium simulations.
It holds $F_\text{th}(x) = -\partial_x \phi_\text{th}(x)$. By construction, $F_\text{th}(x) = 0$ in the AT region.}
\label{forceandpot}
\end{figure}

The reservoirs of the AdResS set-up need to be calibrated to the fluid state they represent.
To this end, a sequence of short equilibrium simulations is needed to self-consistently determine the thermodynamic force $F_\text{th}(x)$ for the pair $(T,\rho)$ of temperature and density of the fluid.
The calculation occurs iteratively, starting from $F_\text{th}(x) = 0$.
The stationary density profile $\rho(x)$ is computed across the AT, $\Delta$, and TR regions for a given form of $F_\text{th}(x)$, which is then updated to reduce gradients of the density (see Ref.~\cite{advtsres}). The iteration ends when the deviation of $\rho(x)$ from the target density is within a prescribed tolerance.
In this work, each iteration step consisted of an MD simulation over $4{,}000\tau$, where the first quarter was skipped in the calculation of $\rho(x)$, allowing the fluid to adjust to the modified value of $F_\text{th}(x)$.
Whereas after about 7 iterations the deviation of the density profile from a constant had dropped below 3\%, we ran about 50 iterations to achieve convergence within 1.5\%.
The resulting functional forms of $F_\text{th}(x)$ and their corresponding potentials $\phi_\text{th}(x)$ are shown in \cref{forceandpot} for the four state points involved in the nonequilibrium AdResS set-ups used here.

\subsection{Validation of the \protect\AdResS{} set-up as a open system}
\label{sec:validation}

\begin{figure*}
\includegraphics[width=.9\linewidth]{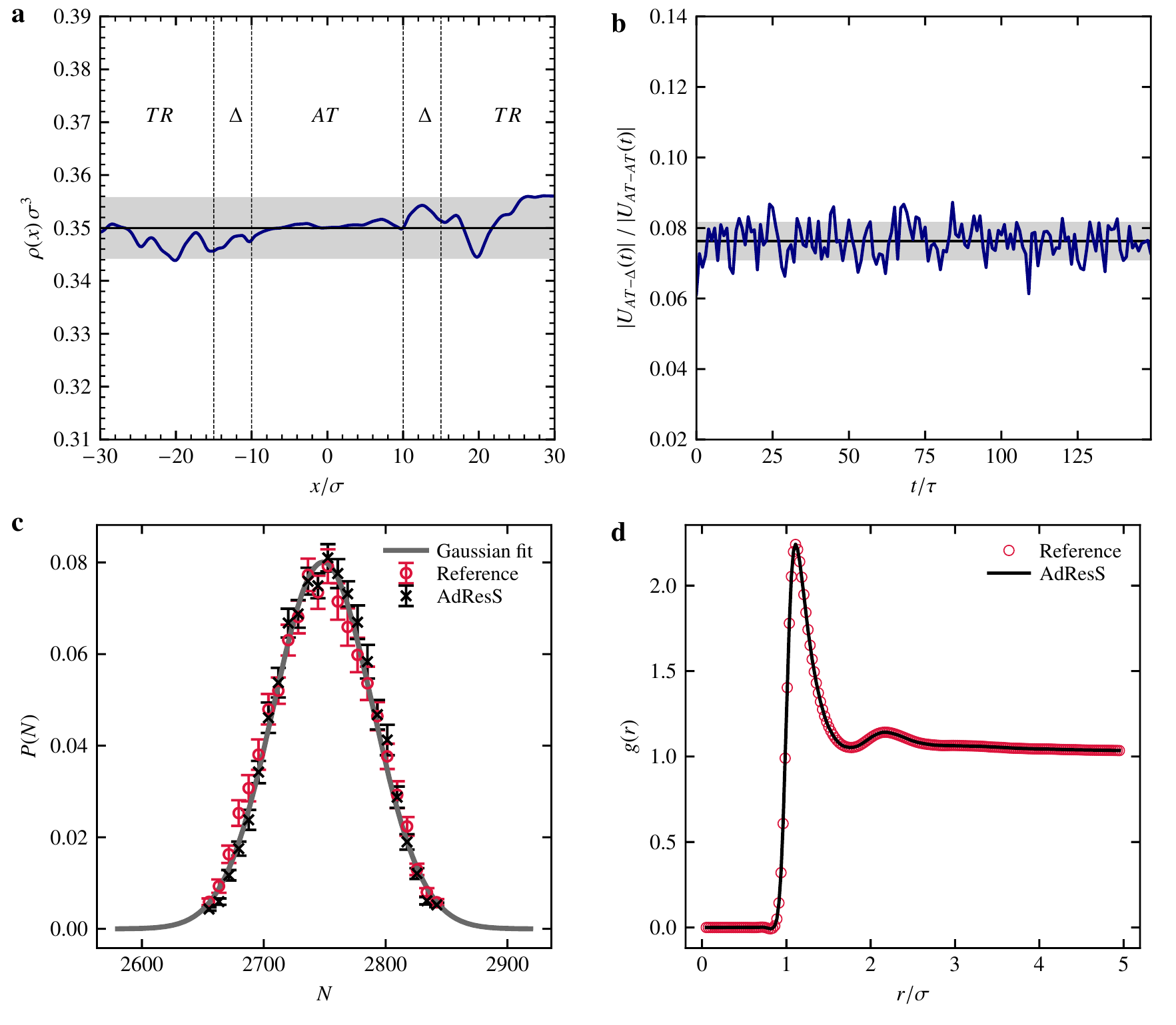}
\caption{Numerical test of the equilibrium AdResS set-up for open-system simulations of a LJ fluid at temperature $T^*=1.40$ and density $\rho^*=0.350$.
Red circles refer to results of the full atomistic reference simulation, where the calculation was restricted to a subdomain that is equivalent to the AT region of the AdResS set-up.
The panels follow the criteria described in the Supplemental section about the validation of the AdResS set up:~
(a)~stationary profile $\rho(x)$ of the number density across the AdResS set-up (blue line) compared to the target density $\rho^*$ (black). The shaded area indicate a tolerance of $\pm 1.5\%$ around $\rho^*$, which was the convergence criterion for the computation of the thermodynamic force.~
(b)~Interaction energy $U_{\text{AT}-\Delta}(t)$ of the AT region of interest with the reservoir relative to the potential energy $U_{\text{AT}-\text{AT}}$ due to the interactions within the AT subsystem, as a function of time. The energy contribution from the reservoir to the AT region is below 8\% on average, with a standard deviation of 0.5\%.~
%
(c)~Probability distribution $P(N)$ of the fluctuating particle number in the AT region of the AdResS set-up (black crosses) compared with results of the reference simulation (red circles). The solid line is a Gaussian fit to the data.~
(d)~Comparison of the radial distribution function $g(r)$ computed from the AT region of the AdResS set-up (black solid line) and from the full atomistic reference; here, the relative deviation is less then 0.3\%.
}
\label{validation}
\end{figure*}

\begin{figure*}[h]
\centering
\includegraphics[width=\linewidth]{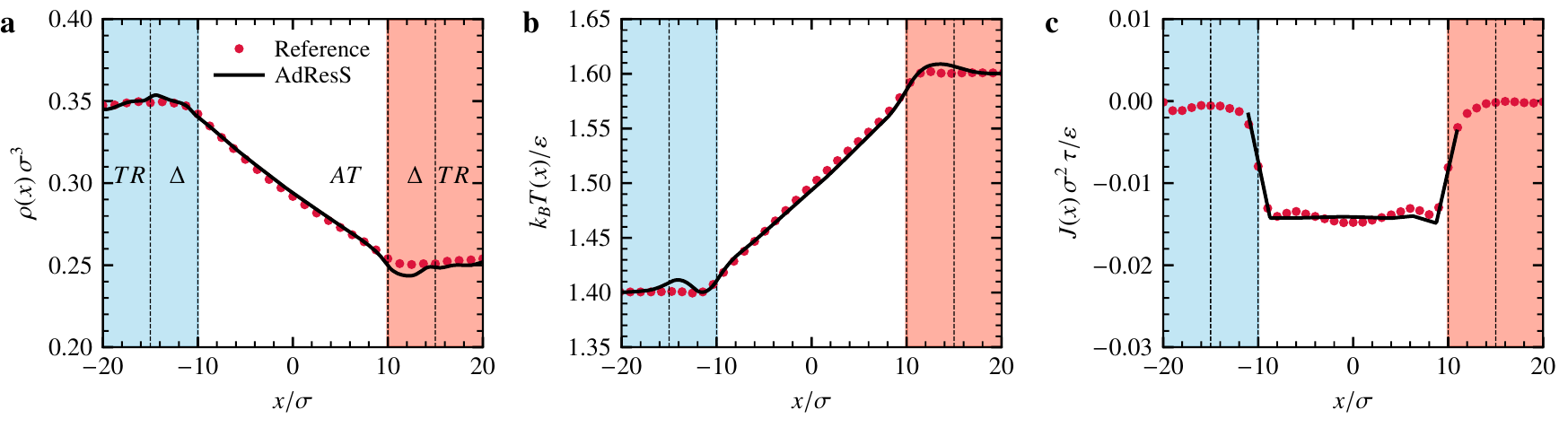}
  \caption{
  Results of the open-system simulation for a supercritical LJ fluid in a thermal gradient obtained from the nonequilibrium AdReS set-up (solid lines) and from a full atomistic simulation serving as reference (dots).
  Only a single AT region is shown, and parts of the tracer regions (TR) have been omitted for clarity.
  The panels display spatial profiles of (a)~the particle number density $\rho(x)$,
  (b)~the temperature $T(x)$, and (c)~the heat flux $J(x)$
  as stationary responses to the nonequilibrium forcing.
  }
\label{dense-gas}
\end{figure*}

Here we report details for the AdResS simulations of the equilibrium fluids and corroborate numerically that the AT region of AdResS is indeed representing a physically well-defined open system.
To this end, the following three conditions must be met by the AT region \cite{physrep,softmatt}.
Data are shown exemplarily for one fluid state only, the results of the other simulations are similar.
\begin{enumerate}[label={(\arabic*)}]
 \item The particle number density and the temperature must be uniform across the AT and $\Delta$ regions and, within a certain tolerance, be equal to their values of the desired thermodynamic state (\cref{validation}a).

 \item The interaction energy $U_{\text{AT}-\Delta}$ of the particles in the AT region with the particles in the $\Delta$ region must be negligible relative to the interaction energy $U_{\text{AT}-\text{AT}}$ amongst the particles in the AT region (\cref{validation}b).

 \item The probability distribution $P(N)$ of the number of particles in the AT region must reproduce the distribution $P(N)$ obtained from an equivalent, open subdomain of the full atomistic reference simulation (\cref{validation}c).
 \end{enumerate}
From (1), we can conclude that, due to the {combined} action of the thermodynamic force {and the thermostat} in the $\Delta$ region, the {AT region} is at the same thermodynamic state point as the reference fluid of a full atomistic simulation.
If (2) is satisfied, there are no sizable energy contributions in the AT region stemming from the reservoir.
This criterion is usually employed in statistical mechanics texts in the definition of {the grand canonical-like} ensemble (see, e.g., \citet{huang}). 
As a consequence of (3), the particle statistics in the AT region is consistent with that of the reference case; in particular, the density (first moment of $P(N)$) and the compressibility (proportional to the variance) are the same.
As a further cross-check, usually automatically fulfilled when (1)--(3) are met, we have tested that the radial distribution function $g(r)$, obtained within the AT region, agrees tightly with the one calculated in the full atomistic simulation (\cref{validation}d).
The quantitative criterion used in this work for the equivalence of the data from the AdResS and the full atomistic reference simulation is a tolerance of $1.5\%$, which is well met by the data shown in \cref{validation}.

\clearpage
\onecolumngrid

\section{Results for the supercritical fluid}

\end{document}